\documentclass{elsart}
\usepackage{amssymb}
\usepackage{amsfonts}
\usepackage{amsmath}

\input{tcilatex}
\begin{document}

%TCIMACRO{\TeXButton{Begin frontmatter}{\begin{frontmatter}}}%
%BeginExpansion
\begin{frontmatter}%
%EndExpansion

%TCIMACRO{%
%\TeXButton{Title}{\title{A New Limit for the Non-Commutative Space-Time Parameter}}}%
%BeginExpansion
\title{A New Limit for the Non-Commutative Space-Time Parameter}%
%EndExpansion

%TCIMACRO{\TeXButton{Author}{\author{Mustafa Moumni}}}%
%BeginExpansion
\author{Mustafa Moumni}%
%EndExpansion
%TCIMACRO{%
%\TeXButton{Address}{\address{Department of Matter Sciences, University Med Khider of Biskra; Algeria}}}%
%BeginExpansion
\address{Department of Matter Sciences, University Med Khider of Biskra; Algeria}%
%EndExpansion
%TCIMACRO{\TeXButton{Address}{\address{m.moumni@univ-batna.dz}}}%
%BeginExpansion
\address{m.moumni@univ-batna.dz}%
%EndExpansion
%TCIMACRO{\TeXButton{Collaborator}{\collab{Achour BenSlama}}}%
%BeginExpansion
\collab{Achour BenSlama}%
%EndExpansion
%TCIMACRO{%
%\TeXButton{Address}{\address{Physics Department, University Mentouri of Constantine; Algeria}}}%
%BeginExpansion
\address{Physics Department, University Mentouri of Constantine; Algeria}%
%EndExpansion
%TCIMACRO{\TeXButton{Address}{\address{a.benslama@yahoo.fr}}}%
%BeginExpansion
\address{a.benslama@yahoo.fr}%
%EndExpansion
%TCIMACRO{\TeXButton{Collaborator}{\collab{Slimane Zaim}}}%
%BeginExpansion
\collab{Slimane Zaim}%
%EndExpansion
%TCIMACRO{%
%\TeXButton{Address}{\address{Physics Department, University Med Abidi of Batna; Algeria}}}%
%BeginExpansion
\address{Physics Department, University Med Abidi of Batna; Algeria}%
%EndExpansion
%TCIMACRO{\TeXButton{Address}{\address{zaim69slimane@yahoo.fr}}}%
%BeginExpansion
\address{zaim69slimane@yahoo.fr}%
%EndExpansion

%TCIMACRO{\TeXButton{Begin abstract}{\begin{abstract}} }%
%BeginExpansion
\begin{abstract}
%EndExpansion
We study space-time noncommutativity applied to the hydrogen atom and the
phenomenological aspects induced. We find that the noncommutative effects
are similar to those obtained by considering the extended charged nature of
the proton in the atom. To the first order in the noncommutative parameter,
it is equivalent to an electron in the fields of a Coulomb potential and an
electric dipole and this allows us to get a bound for the parameter. In a
second step, we compute noncommutative corrections of the energy levels and
find that they are at the second order in the parameter of noncommutativity.
By comparing our results to those obtained from experimental spectroscopy,
we get another limit for the parameter.%
%TCIMACRO{\TeXButton{End abstract}{\end{abstract}}}%
%BeginExpansion
\end{abstract}%
%EndExpansion

%TCIMACRO{\TeXButton{Begin keywords}{\begin{keyword}} }%
%BeginExpansion
\begin{keyword}
%EndExpansion
Noncommutative space-time, Hydrogen atom, Dipole potential 
%TCIMACRO{\TeXButton{End keyword}{\end{keyword}}}%
%BeginExpansion
\end{keyword}%
%EndExpansion

%TCIMACRO{\TeXButton{End frontmatter}{\end{frontmatter}}}%
%BeginExpansion
\end{frontmatter}%
%EndExpansion

\textit{MSC\ codes}: 81T75 , 81Q05 , 81V45

\textit{Subject Classification}: Quantum Mechanics

\section{Introduction:}

In the recent years there has been a large interest in the study of
noncommutative geometry. The idea of taking space-time coordinates to be
noncommutative goes back to the thirties of the last century. The goal was
that the introduction of a noncommutative structure to space-time at small
length scales could introduce an effective cut off which regularize
divergences in quantum field theory. However this theory was plagued with
several problems such as the violation of unitarity and causality, which
make people abandon it. However noncommutative geometry was pursued on the
mathematical side and especially with the work of Connes in the eighties of
the last century [1].

In 1999, the interest for noncommutative geometry is renewed by the work of
Seiberg and Witten on string theory [2]. They showed that the dynamics of
the endpoints of an open string on a D-brane in the presence of a magnetic
background field can be described by a Yang-Mills theory on a noncommutative
space-time.

Noncommutative space-time is a deformation of the ordinary one in which the
coordinates are promoted to Hermitian operators which do not commute:%
\begin{equation}
\left[ x_{nc}^{\mu },x_{nc}^{\nu }\right] =i\theta ^{\mu \nu }=iC^{\mu \nu
}/\left( \Lambda _{nc}\right) ^{2};\mu ,\nu =0,1,2,3  \label{eqn1}
\end{equation}%
where $\theta ^{%
%TCIMACRO{\U{b5}}%
%BeginExpansion
{\mu}%
%EndExpansion
\nu }$\ is a deformation parameter and $nc$\ indices denote noncommutative
coordinates. Ordinary space-time is obtained by making the limit $\theta ^{%
%TCIMACRO{\U{b5}}%
%BeginExpansion
{\mu}%
%EndExpansion
\nu }\rightarrow 0$. The noncommutative parameter is an anti-symmetric real
matrix, assumed here to be constant. In natural units where $\hbar =1$ and $%
c=1$, $\theta ^{%
%TCIMACRO{\U{b5}}%
%BeginExpansion
{\mu}%
%EndExpansion
\nu }$ is in $eV^{-2}$, $\Lambda _{nc}$ is the energy scale where the
noncommutative effects of the space-time will be relevant and $C^{%
%TCIMACRO{\U{b5}}%
%BeginExpansion
{\mu}%
%EndExpansion
\nu }$\ are dimensionless parameters (Otherwise, we have to distinguish
between the space-space case of noncommutativity where $\theta ^{%
%TCIMACRO{\U{b5}}%
%BeginExpansion
{\mu}%
%EndExpansion
\nu }$ is in $m^{2}$ and the space-time case where the unit of $\theta ^{%
%TCIMACRO{\U{b5}}%
%BeginExpansion
{\mu}%
%EndExpansion
\nu }$\ is $ms$). For a review, one can see reference [3].

In the literature, there are a lot of phenomenological studies giving bounds
on the noncommutative parameter. For example, the OPAL collaboration founds $%
\Lambda _{nc}\geq 140~GeV$ [4], various noncommutative QED processes give
the range $\Lambda _{nc}\geq 500~GeV-1.7~TeV$ [5], high precision atomic
experiment on the Lamb shift in the hydrogen atom gives the limit $\Lambda
_{nc}\leq 10^{4}~GeV$\ [6] but this bound was corrected in [7] to $\Lambda
_{nc}\leq 6~GeV$; all these bounds deal with space-space noncommutativity.
For the space-time case, the bound $\theta \lesssim 9.51\times 10^{-18}m.s$
was found from quantum gravity considerations [8], the bound $\theta
_{st}\lesssim (0.6GeV)^{-2}$ was determined in [9] from theoretical limit of
the Lamb shift in H atom. Using the loss of Lorentz invariance in the
theory, some specific models gives the bound $\Lambda _{nc}\geq 10~TeV$ from
CMB data [10] or the bound $\Lambda _{nc}\gtrsim 10^{16}~GeV$\ from particle
phenomenology [11] (Recently, many works study Lorentz invariant
interpretation of the theory [12-14] and the references therein). A well
documented review on noncommutative parameter bounds can be found in [15].

We work here on the space-time version of the noncommutativity; thus instead
of (1), we use:%
\begin{equation}
\left[ x_{st}^{j},x_{st}^{0}\right] =i\theta ^{j0};j=1,2,3  \label{eqn2}
\end{equation}

The $st$ indices are for space-time noncommutative coordinates. We are
interested in the phenomenological consequences and we focus on the hydrogen
atom because it is a well studied quantum system and so it can be taken as
an excellent test for noncommutative signatures. We start by writing the Schr%
\"{o}dinger equation for the H-atom in the noncommutative space-time. We get
a first limit for the parameter from the fact that the solutions must be
real. In a second step, we compute the corrections of the energy levels
induced by noncommutativity and find another limit for the parameter by
comparing to results coming from high precision spectroscopy.

The aim of this work is to find an upper limit for the noncommutative
parameter in order to have a bound spectrum of the hydrogen atom.

\section{Noncommutative Hydrogen Atom:}

To build the Hamiltonian and as solutions to the relations (2), we choose
the transformations: 
\begin{equation}
x_{st}^{j}=x^{j}+i\theta ^{j0}\partial _{0}  \label{eqn3}
\end{equation}%
All the other coordinates remain as usual. The usual space coordinates $%
x^{j} $\ satisfy the usual canonical permutation relations. For convenience
we use the vectorial notation: 
\begin{equation}
\overrightarrow{r}_{st}=\overrightarrow{r}+i\overrightarrow{\theta }\partial
_{0}=\overrightarrow{r}-\overrightarrow{\theta }E/\hbar  \label{eqn4}
\end{equation}%
We have used the fact that $i\partial _{0}\psi =\left( E/\hbar \right) \psi $
and the notation:%
\begin{equation}
\overrightarrow{\theta }\equiv \left( \theta ^{10},\theta ^{20},\theta
^{30}\right) =\left( \theta ^{1},\theta ^{2},\theta ^{3}\right)  \label{eqn5}
\end{equation}%
The relations (3) and (4) can be seen as a Bopp's shift [16].

We are dealing with the stationary Schr\"{o}dinger equation, and this allows
us to consider the energy as a constant parameter. In our computation, we
use the standard equation; it is possible because our choice in the
transformations (3) leaves the coordinate $x^{0}$ and all the momentums $%
p^{\mu }$ unchanged and also, it was shown in [17] and [18] that "the
spectrum is unchanged if one replaces the standard Schr\"{o}dinger equation
with its noncommutative image if the spatial coordinates commute in its
noncommutative form (the only noncommutativity being between time and space
coordinates)". The kinetic energy does not change since it depends on the
momentum which remains unchanged; so we look for the Coulomb potential and
construct its noncommutative image. We write it as the usual one but with
the new coordinates: 
\begin{equation}
V_{nc}(r)=-e^{2}/r_{st}=-e^{2}\left( \tsum x_{st}^{j}\cdot x_{st}^{j}\right)
^{-1/2}  \label{eqn6}
\end{equation}%
or: 
\begin{equation}
V_{nc}(r)=-e^{2}\left[ (\overrightarrow{r}-\overrightarrow{\theta }E/\hbar
)^{2}\right] ^{-1/2}  \label{eqn7}
\end{equation}%
We write:%
\begin{equation}
\frac{1}{r_{st}}=\frac{1}{r}\left( 1-2\frac{E}{\hbar }\frac{\overrightarrow{r%
}\cdot \overrightarrow{\theta }}{r^{2}}+\frac{E^{2}}{\hbar ^{2}}\frac{\theta
^{2}}{r^{2}}\right) ^{-1/2}  \label{eqn8}
\end{equation}%
To make the development in series of the expression, we choose:%
\begin{equation}
\varepsilon =-2\left( E/\hbar \right) (\overrightarrow{r}\cdot 
\overrightarrow{\theta })/r^{2}+\left( E/\hbar \right) ^{2}\theta ^{2}/r^{2}
\label{eqn9}
\end{equation}%
and consider it as a small parameter because of the smallness of $\theta $.
For example, to the second order of $\varepsilon $, one has:%
\begin{equation}
\left( 1+\varepsilon \right) ^{-\frac{1}{2}}=1-\varepsilon /2+3\varepsilon
^{2}/8+O\left( \varepsilon ^{3}\right)  \label{eqn10}
\end{equation}%
Using (9), we find:%
\begin{equation}
\frac{1}{r_{st}}=\frac{1}{r}\left[ 1+\frac{E}{\hbar }\frac{\overrightarrow{r}%
\cdot \overrightarrow{\theta }}{r^{2}}-\frac{E^{2}}{2\hbar ^{2}}\frac{\theta
^{2}}{r^{2}}+\frac{3E^{2}}{2\hbar ^{2}}\frac{(\overrightarrow{r}\cdot 
\overrightarrow{\theta })^{2}}{r^{4}}+O(\theta ^{3})\right]  \label{eqn11}
\end{equation}%
The other higher order terms are higher power in $\theta $ and can be
neglected (We restrict ourselves to the second order). With an adequate
choice of the coordinates:%
\begin{equation}
\overrightarrow{\theta }=\theta ^{30}\overrightarrow{k}=\theta _{st}%
\overrightarrow{k}  \label{eqn12}
\end{equation}%
the noncommutative Coulomb potential writes ($\vartheta $\ represents the
azimuthal angle):%
\begin{equation}
V_{nc}(r)=-\frac{e^{2}}{r}-\frac{e^{2}E\theta _{st}}{\hbar }\frac{\cos
\vartheta }{r^{2}}-\frac{e^{2}E^{2}\theta _{st}^{2}}{2\hbar ^{2}}\frac{%
(3\cos ^{2}\vartheta -1)}{r^{3}}+O(\theta ^{3})  \label{eqn13}
\end{equation}%
This expression is similar to the potential energy of an electric charge $q$
due to the presence of a distribution of charges $q_{i}$: 
\begin{subequations}
\begin{gather}
V(r)=\tsum (q/4\pi \epsilon _{0})q_{i}/r_{i}  \label{eqn14} \\
V(r)=\frac{q}{4\pi \epsilon _{0}}\left[ \frac{\sum q_{i}}{r}+\frac{\sum
a_{i}q_{i}\cos \vartheta _{i}}{r^{2}}+\frac{\sum a_{i}^{2}q_{i}\left( 3\cos
^{2}\vartheta _{i}-1\right) }{2r^{3}}+O(a_{i}^{3})\right]
\end{gather}%
where: 
\end{subequations}
\begin{equation}
\overrightarrow{r_{i}}=\overrightarrow{A_{i}M}=\overrightarrow{OM}-%
\overrightarrow{OA_{i}}=\overrightarrow{r}-\overrightarrow{a_{i}}
\label{eqn15}
\end{equation}%
Here, $M$ and $A_{i}$ refer to the positions of the charges $q$ and $q_{i}$
respectively and $O$\ is the origin.

The relations (13) and (14b) can be compared term by term; they are the
Coulomb term and both the dipolar and the quadrupolar contributions to the
potential. This implies that the noncommutative Coulomb potential is
equivalent to an electron in a field of a distribution of positive and
negative charges which are not equal ($\tsum q_{i}\neq 0$ and a positive net
charge here) so it gives the Coulomb contribution, and the distribution is
not spherically\ symmetric and this adds the multipolar contributions. Such
a distribution exists in the hydrogen atom in the proton; it is a extended
positively charged system composed of three quarks, two have positive
charges and the third has a negative one.

As mentioned in [19], due to the fact that the proton has a structure and is
a composite particle, noncommutativity cannot be applied to it as for
elementary particles like electron, and the proton behaves essentially as a
commutative particle in the noncommutative hydrogen atom. Thus, we applied
noncommutativity only to the electron; however we found that this is
equivalent to consider the internal electric structure of the proton. As the
relations (4) and (15) are similar, the equivalence is true to any order of
the development.

To find limits on $\theta $, we consider the first order terms; this gives
the potential:%
\begin{equation}
V_{nc}(r)=-e^{2}/r-e^{2}\left( E/\hbar \right) (\overrightarrow{r}\cdot 
\overrightarrow{\theta })/r^{3}  \label{eqn16}
\end{equation}%
The additional term to the Coulomb one is similar to an electric dipole
potential:%
\begin{equation}
V_{ed}\left( r\right) =e(\overrightarrow{D}\cdot \overrightarrow{r})/r^{3}
\label{eqn17}
\end{equation}%
where the dipole moment is proportional to the noncommutative\ parameter:%
\begin{equation}
\overrightarrow{D}=eE\overrightarrow{\theta }/\hbar
\end{equation}%
Using the angular notation, the potential becomes:%
\begin{equation}
V_{nc}(r)=-e^{2}/r-eD\left( \cos \vartheta \right) /r^{2};D=eE\theta
_{st}/\hbar  \label{eqn19}
\end{equation}%
The two terms do not correspond to a potential of an usual electric dipole
with zero net charge (a pure electric dipole). The expression looks like the
potential of a molecular anion; a system composed with a molecule and an
electron (The molecule is generally taken as a pure dipole in chemistry). If
the molecule is not neutral, the system is capable of supporting bound
states if the dipole moment is smaller than a critical value which depends
on the azimuthal quantum number [20].%
\begin{equation}
D\leq D_{m}  \label{eqn20}
\end{equation}%
We have $\sum q_{i}=+\left\vert q_{e}\right\vert $ in our case and since
there is no limitation on the dimensions of the anion, we apply this result
to our system. As the smallest value for this limit is when the quantum
number vanishes and because the s-orbital exists in the hydrogen atom, we
choose this value as a limit for our computation:%
\begin{equation}
D\leq D_{0}=5.421\cdot 10^{-30}Cm\ (=0.6393a.u)  \label{eqn21}
\end{equation}%
From (19),(20) and (21), we have:%
\begin{equation}
\theta _{st}\leq \hbar D_{0}/eE  \label{eqn22}
\end{equation}%
and to consider all the states of the electron, we take the smallest value
which corresponds to the first energy state. So we find the fundamental
limit:%
\begin{equation}
\theta _{st}\lesssim 1.6\cdot 10^{-27}ms\approx (0.3keV)^{-2}  \label{eqn23}
\end{equation}%
The limit is $10^{9}$\ times smaller\ than the one obtained using quantum
gravity considerations [8].

We can get another limit for $\theta $ by looking to corrections of energy
levels and comparing them to the experimental results in spectroscopy. We
must first note that the term $\left( r^{-2}\cos \vartheta \right) $ in the
potential, gives non-zero matrix elements between states with $\left( \Delta
n\neq 0;\Delta l=\pm 1;\Delta m=0\right) $; thus we have no diagonal terms
and hence no corrections to the 1st order in $\theta $. So the term of order
1 in $\theta $ in the potential gives corrections of order 2 in $\theta $
for energy. This is why we must also consider corrections of energy in the
1st order in $\theta ^{2}$ which are diagonal elements coming from the term
of order 2 in $\theta $ in the potential.

We write the potential as:%
\begin{equation}
V_{nc}(r)=-e^{2}\left[ \frac{1}{r}+\left( \frac{E\theta }{\hbar }\right) 
\frac{\cos \vartheta }{r^{2}}+\left( \frac{E\theta }{\hbar }\right) ^{2}%
\frac{3\cos ^{2}\vartheta -1}{2r^{3}}+O(\Theta ^{3})\right]  \label{eqn24}
\end{equation}%
and use a new parameter $\Theta =E\theta /\hbar $\ to\ simplify the
expression: 
\begin{equation}
V_{nc}(r)=-e^{2}\left[ \frac{1}{r}+\Theta \frac{\cos \vartheta }{r^{2}}%
+\Theta ^{2}\frac{3\cos ^{2}\vartheta -1}{2r^{3}}+O(\Theta ^{3})\right]
\label{eqn25}
\end{equation}%
So we must consider all matrix elements of the term $\left( r^{-2}\cos
\vartheta \right) $ and the diagonal ones of the term $\left( 2(3\cos
^{2}\vartheta -1)r^{-3}\right) $ and then calculate the eigenvalues of the
matrix obtained. For example, if we consider levels from $n=1$\ to $n=3$\
and use the base:%
\begin{equation}
\left( \left( 1,0,0\right) ,\left( 2,0,0\right) ,\left( 2,1,0\right) ,\left(
3,0,0\right) ,\left( 3,1,0\right) ,\left( 3,2,0\right) \right)  \label{eqn26}
\end{equation}%
where the parentheses mean the state $\left( n,l,m\right) $, we find the
matrix:%
\begin{equation}
\left( 
\begin{array}{cccccc}
\frac{-1}{2a} & 0 & \frac{-2\sqrt{2}\Theta }{27a^{2}} & 0 & \frac{-\Theta }{%
12\sqrt{2}a^{2}} & 0 \\ 
0 & \frac{-1}{8a} & 0 & 0 & \frac{-16\Theta }{1875a^{2}} & 0 \\ 
\frac{-2\sqrt{2}\Theta }{27a^{2}} & 0 & \frac{-1}{8a}+\frac{\Theta ^{2}}{%
60a^{3}} & \frac{-4\sqrt{2}\Theta }{625\sqrt{3}a^{2}} & 0 & \frac{-64\Theta 
}{3125\sqrt{3}a^{2}} \\ 
0 & 0 & \frac{-4\sqrt{2}\Theta }{625\sqrt{3}a^{2}} & \frac{-1}{18a} & 0 & 0
\\ 
\frac{-\Theta }{12\sqrt{2}a^{2}} & \frac{-16\Theta }{1875a^{2}} & 0 & 0 & 
\frac{-1}{18a}+\frac{2\Theta ^{2}}{405a^{3}} & 0 \\ 
0 & 0 & \frac{-64\Theta }{3125\sqrt{3}a^{2}} & 0 & 0 & \frac{-1}{18a}+\frac{%
2\Theta ^{2}}{2835a^{3}}%
\end{array}%
\right)  \label{eqn27}
\end{equation}%
Here $a$ represents the Bohr radius.

We will take as test levels $1S$ and $2S$ because we have the best
experimental precision for the transition between them [21]:%
\begin{equation}
f_{1S-2S}=\left( 2446061102474851\pm 34\right) Hz  \label{eqn28}
\end{equation}%
In our computation, we consider the states from $\left( 1,0,0\right) $ to $%
\left( 5,3,0\right) $ because the contributions from the higher ones to
those considered above ($1S$ and $2S$) are negligible (We have made
calculations up to the level $n=11$ without noticing changes in the results
because the matrix elements decreases with the value of $\triangle n$).

For these two levels, we obtain the noncommutative corrections: 
\begin{subequations}
\begin{gather}
\Delta E_{1S}=-0.04187\left( E\theta _{st}/\hbar \right) ^{2}e^{2}/a^{3} \\
\Delta E_{2S}=-0.00173\left( E\theta _{st}/\hbar \right) ^{2}e^{2}/a^{3}
\end{gather}%
and for the transition, this gives the correction: 
\end{subequations}
\begin{equation}
\Delta E_{nc}\left( 1S-2S\right) =0.04014\left( E\theta _{st}/\hbar \right)
^{2}e^{2}/a^{3}  \label{eqn30}
\end{equation}%
Comparing with the precision of the experimental value in (28), we obtain:%
\begin{equation}
\theta _{st}\lesssim (0.3MeV)^{-2}  \label{eqn31}
\end{equation}%
This is a significant improvement of the limit obtained previously ($10^{6}$%
\ times better).

In [9], the author obtained the limit: $\theta _{st}\lesssim (0.6GeV)^{-2}$
which is smaller than ours. The method used in this work is to consider the
matrix element between two states as the correction of the transition's
spectrum between the two corresponding levels without calculating the
energies. This gives a correction at the 1st order in $\theta $, although
the matrix elements are not diagonal in his case also, whereas the
corrections are at the 2nd order in our case. To compare with this work, we
can use the same method. We cannot use the Lamb shift because transitions
with $\left( \triangle n=0\right) $ are forbidden in our case if we consider
only the term of order 1 in $\theta $\ in the potential. Therefore, we use
the $1S-3P$ transition because the experimental precision is good enough in
this case; the accuracy is $1.44kHz$ [22].

We have:%
\begin{equation}
e^{2}\left( E/\hbar \right) \theta _{st}\left\langle 100\left\vert
r^{-2}\cos \vartheta \right\vert 310\right\rangle =e^{2}\left( E/\hbar
\right) \theta _{st}/(12\sqrt{2}a)  \label{eqn32}
\end{equation}

And by comparing to the experimental accuracy, we get:%
\begin{equation}
\theta _{st}\lesssim (0.16GeV)^{-2}  \label{eqn33}
\end{equation}%
It is a little larger than the limit in [9] but it is in the same range. It
must be noted that this is mainly due to the fact that $P$-states have a
short natural lifetime and because of that, their spectral lines are rather
broad and the accuracy in our case is weak compared to that of [9] where the
Lamb shift is used.

\section{Conclusion:}

In this work, we study the hydrogen atom in the context of\ noncommutative
space-time and specially the phenomenological effects resulting. We found
that applying noncommutativity to the electron of the atom is equivalent to
consider the extended nature of the proton in the nucleus; not as a QCD
particle but as a composite electrically charged system. By making the
similitude with molecular anions, we got an upper limit for the
noncommutative parameter ($\theta _{st}\lesssim (0.3keV)^{-2}$) which is
smaller than the one obtained by quantum gravity considerations (The
molecular anion is considered here to have both a net charge and a dipole
moment). In [20], it is shown that if the dipole moment exceeds the critical
value considered here, then the reality of the representation will be
violated\ in the angular\ part of the solution of the Schr\"{o}dinger
equation. It would be interesting to see what would happen to the orbital
near this limit using tools like Poincare sections. The limit is fundamental
because if the noncommutative\ parameter exceeds the value obtained here,
the fundamental state of the hydrogen atom ($n=1;l=0;m=0$) no longer exists
and we consider this limit as a critical value for the noncommutative
parameter $\theta $.

In a second step, we calculated the noncommutative corrections to energy
levels. We found that they are in the second order in $\theta $ because the
1st order term in $\theta $ in the noncommutative Coulomb potential has off
diagonal contributions to the Hamiltonian matrix. By comparing to results
from high precision spectroscopy, it gave us a new limit ($\theta
_{st}\lesssim (0.3MeV)^{-2}$) which improves the first one in a significant
way. The signature on the energy levels of the H-atom in our case is
different from the one obtained in the space-space case in\ [6] where the
additional terms are diagonal and the corrections were in the 1st order in $%
\theta $.

In [17] and [18], the authors found that the hydrogen spectrum does not
change if there is only space-time noncommutativity. Their result is based
on the assumption that the Hamiltonian does not contain a $P_{0}=i\partial
_{0}$ dependence; the invoked reason is the absence of such a dependence in
the commutative case where $\theta =0$. In our calculation, the Hamiltonian
does contain the $P_{0}=i\partial _{0}$\ term in its potential part as one
can see from\ (7) and this is due to the solutions (3) of the commutation
relations (2) (This choice is done \`{a} la Chaichian [6] and as we
discovered recently, it is similar to the work done in [23]). This $P_{0}$
term is multiplied by the parameter $\theta $\ and thus it disappears when $%
\theta \rightarrow 0$ and again we have no $P_{0}$ dependence in the
commutative case. By acting on the wave function in the Schr\"{o}dinger
equation, this $P_{0}$\ term is the cause of the energy shift through
changes in the expression of the potential in the Hamiltonian. We can
consider our work as a generalization of [17] when there is such $P_{0}$
dependence in the Hamiltonian.

\textbf{Acknowledgments:}

Mr Moumni M would like to thanks Mr Delenda Y, Mr Aouachria M and especially
Mr Ydri B and Mr Bouchareb A for their discussions and recommendations and
also Mrs Merzougui G for her encouragements.


\begin{thebibliography}{99}
\bibitem{bibitem1} A. Connes, Noncommutative Geometry,\ Academic Press, San
Diego, CA, 1994.

\bibitem{bibitem2} N. Seiberg, E. Witten, String theory and Noncommutative
Geometry, JHEP 09 (1999) 032.

\bibitem{bibitem3} R.J. Szabo, Quantum Field Theory on Noncommutative
Spaces, Phys.Rep. 378 (2003) 207-299.

\bibitem{bibitem4} G. Abbiendi et al, Test of Non-Commutative QED Effect in
the Process $e^{+}e^{-}\rightarrow \gamma \gamma $,\ Phys.Lett. B568 (2003)
181-190.

\bibitem{bibitem5} J.L. Hewett, F.J. Petriello, T.G. Rizzol, Signals for
Non-Commutative Interactions at Linear Colliders, Phys.Rev. D64 (2001)
075012.

\bibitem{bibitem6} M. Chaichian, M.M. Sheikh-Jabbari, A. Tureanu, Hydrogen
Atom Spectrum and the Lamb Shift in Noncommutative QED, Phys.Rev.Lett. 86
(2001) 2716-2719.

\bibitem{bibitem7} A. Stern, Noncommutative Point Sources, Phys.Rev.Lett.
100 (2008) 061601.

\bibitem{bibitem8} A. Saha, Time-Space Noncommutativity in Gravitational
Quantum Well scenario, Eur.Phys.J. C51 (2007) 199-205 (and the references
therein).

\bibitem{bibitem9} A. Stern, Particle-like Solutions to Classical
Noncommutative Gauge Theory, Phys.Rev. D78 (2008) 065006.

\bibitem{bibitem10} E. Akofor, A.P. Balachandran, A. Joseph, L. Pekowsky,
B.A. Qureshi, Constraints from CMB on Spacetime Noncommutativity and
Causality Violation, Phys.Rev. D79 (2009) 063004.

\bibitem{bibitem11} A. Joseph, Particle Phenomenology on Noncommutative
Spacetime, Phys.Rev. D79 (2009) 096004.

\bibitem{bibitem12} M. Chaichian, P. Kulish, K. Nishijima, A. Tureanu, On a
Lorentz-Invariant Interpretation of Noncommutative Space-Time and Its
Implications on Noncommutative QFT, Phys.Lett. B604 (2004) 98-102.

\bibitem{bibitem13} C.D. Carone, H.J. Kwee, Unusual High-Energy
Phenomenology of Lorentz-Invariant Noncommutative Field Theories, Phys.Rev.
D73 (2006) 096005.

\bibitem{bibitem14} S. Saxell, On General Properties of Lorentz Invariant
Formulation of Noncommutative Quantum Field Theory, Phys.Lett. B666 (2008)
486-490.

\bibitem{bibitem15} R.J. Szabo, Quantum Gravity, Field Theory and Signatures
of Noncommutative Spacetime, Gen.Relativ.Gravit. 42 (2010) 1-29.

\bibitem{bibitem16} S. Dulat, K. Li, The Aharonov-Casher Effect for Spin-1
Particles in Non-Commutative Quantum Mechanics, Eur.Phys.J. C54 (2008)
333-337.

\bibitem{bibitem17} A.P. Balachandran, T.R. Govindarajan, C. Molina, P.
Teotonio-Sobrinho, Unitary Quantum Physics with Time-Space Noncommutativity,
JHEP 0410 (2004) 072.

\bibitem{bibitem18} A.P. Balachandran, A. Pinzul, On Time-Space
Noncommutativity for Transition Processes and Noncommutative Symmetries,
Mod.Phys.Lett. A20 (2005) 2023-2034.

\bibitem{bibitem19} M. Chaichian, M.M. Sheikh-Jabbari, A. Tureanu, Comments
on the Hydrogen Atom Spectrum in the Noncommutative Space, Eur.Phys.J. C36
(2004) 251-252.

\bibitem{bibitem20} A.D. AlHaidari, Analytic Solution of the Schr\"{o}dinger
Equation for an Electron in the Field of a Molecule with an Electric Dipole
Moment,\ Ann.Phys. 323 (2008) 1709-1728.

\bibitem{bibitem21} T.W. Hansch et al, Precision Spectroscopy of Hydrogen
and Femtosecond Laser Frequency Combs, Philos.Trans.R.Soc. A363 (2005)
2155--2163.

\bibitem{bibitem22} National Istitute of Standards and Technology,
http://www.nist.gov/physlab/data/hdel/

\bibitem{bibitem23} L. Dabrowski, P. Parashar, \ A Free Particle in
Noncommutative Space-Time, Czech.J.Phys. 46 (1996) 1211-1215
(SISSA-134-95-FM).
\end{thebibliography}
\end{document}